\documentclass[conference]{IEEEtran}

\usepackage{times}
\usepackage{epsfig}
\usepackage{amsmath}
\usepackage{amsfonts}
\usepackage{graphicx}
\usepackage{amssymb}
\usepackage{amstext}
\usepackage{latexsym}
\usepackage{color}
\usepackage{ifthen}
\usepackage{multirow}
\usepackage{verbatim}
\usepackage{array,tabularx}
\usepackage[mathscr]{euscript}
\usepackage{tikz}
\usepackage{accents}
 \usepackage{cite}

\newcommand{\be}{\begin{equation}}
\newcommand{\ee}{\end{equation}}
\newcommand{\bear}{\begin{eqnarray}}
\newcommand{\eear}{\end{eqnarray}}
\newcommand{\bears}{\begin{eqnarray*}}
\newcommand{\eears}{\end{eqnarray*}}
\newcommand{\bi}{\begin{itemize}}
\newcommand{\ei}{\end{itemize}}
\newcommand{\ben}{\begin{enumerate}}
\newcommand{\een}{\end{enumerate}}

\newtheorem{theorem}{Theorem}

\newcommand{\ubar}[1]{\underaccent{\bar}{#1}}

\IEEEoverridecommandlockouts

\renewcommand{\baselinestretch}{1}

\begin{document}
\title{Securing Data against Limited-Knowledge Adversaries in Distributed Storage Systems}
\author{
\IEEEauthorblockN{Rawad Bitar,  Salim El Rouayheb\\ ECE Department\\ IIT, Chicago}

}
\maketitle

\begin{abstract}
We study the problem of constructing secure regenerating codes that protect data integrity in distributed storage systems (DSS) in which some nodes may be compromised by a malicious  adversary. The adversary can corrupt the data stored on  and transmitted by  the nodes under its control. The ``damage" incurred by the actions of the adversary depends on how much information it knows about the data in the whole DSS. We focus on the limited-knowledge model in which the adversary knows only the data on the nodes under its control. The only secure capacity-achieving codes known in the literature for this model are for the bandwidth-limited regime and repair degree $d=n-1$, i.e.,  when a node fails in a  DSS with $n$ nodes all the remaining $n-1$ nodes are contacted for repair. We extend these results to the more general case of $d\leq n-1$ in the bandwidth-limited regime. Our capacity-achieving scheme is based on the use of product-matrix codes with special hashing functions and allow the identification of the compromised nodes and their elimination from the DSS while preserving the data integrity.

\end{abstract}
\section{Introduction}\label{sec:Intro}
We consider the problem of securing data in distributed storage systems (DSS) under failure and  repair (rebuilding) dynamics against a malicious adversary that can control a certain number of nodes in the system.
DSS experience frequent node failures due to the use of inexpensive commodity hardware \cite{GFS,schroeder2007disk}. Data redundancy is used to prevent from data loss. Typically, replication codes are used and multiple copies of the data, typically $3$,  are stored in the DSS. Recently, major cloud storage companies \cite{LRCMicrosoft, XorbaFacebook}  have started using   erasure codes, such as regenerating codes \cite{DGWWR07} and locally repairable codes \cite{LRCMicrosoft},  to achieve data reliability with a  lower storage cost and better tradeoffs with other system resources, such as repair bandwidth and data locality.

We assume that there is an adversary that  can corrupt the data stored on the nodes under its control and all of their outgoing messages. The damage incurred by the adversary depends on how much information it knows about the stored file. We focus on the {\em limited-knowledge} model \cite{PRK11, JLKHKME08} in which the only information the adversary knows about the  stored data    comes from reading the data on the nodes  under its control. This is in contrast with an  {\em omniscient} adversary that  although it  controls a limited number of nodes, it has complete knowledge of all the data in  the DSS. Due to the distributed nature of DSS, a limited-knowledge adversary may be a more suitable model in many applications.

Classical  codes, such as Reed-Solomon codes, can be used to correct errors. However, they may not be well suited to the secure distributed storage problem under consideration, for two main reasons: (1) they are designed for an omniscient adversary or random errors, (2) they result in high repair bandwidth. Take for instance a  replication code that stores $4$ copies of a file on $4$ different storage nodes, one of which is controlled by an adversary. If one of the other nodes fails and is repaired by contacting the remaining $3$ nodes, it has to download all their data in order to do majority decoding and regenerate an uncorrupted copy of the file. This corresponds to a repair bandwidth equal to $3$ times the size of the stored file. To address these challenges, we  construct   regenerating codes that can secure the data and
enable an efficient repair process.

Our objective is to achieve {\em information theoretic security} of the data that guarantees security even if the adversary has unlimited computation power. {\em Computational security} can be possibly achieved by storing on a trusted server cryptographic  hash functions (SHA-2, etc.)  of the data on each node. Then, the hashes of the downloaded data can be computed and compared to the trusted hashes to determine possible corruption.  This would work given the empirical assumption that it is computationally hard for the adversary to create a hash collision. Here, we do not make this assumption  and  show that there is no extra cost for achieving information theoretic security  (as opposed to the omniscient case). Applications may include long term storage \cite{storer2008potshards}, as hash functions that are hard to break now are  more likely to be broken in the  future.

\noindent\paragraph*{Related work} The problem of studying the information theoretic security of  DSS with repair dynamics was first studied in \cite{PRK10,PRKISIT11, PRK11}, where different adversarial models were considered: passive (eavesdropper), active  (omniscient vs. limited-knowledge). Further results on  achieving security in the DSS against eavesdropping appeared in \cite{RSKGlobecom11, RKSV13, Tandon, KS14} and omniscient adversary in \cite{RSK12,silberstein2012error}. Schemes for protecting  data in DSS against random errors were studied in  \cite{DDH10,NCAudit}.

\noindent\paragraph*{Contribution} An upper bound on the secure capacity of the DSS, also called resiliency capacity,  for  the limited-knowledge model was proved in \cite{PRK11} (see Eq.~\eqref{eq:UpperBound}), as well as a scheme that achieves this upper bound  in the bandwidth-limited regime. This scheme is restricted to a repair degree $d=n-1$, i.e., all remaining nodes are contacted to repair a failed node. In this paper, we prove that the upper bound in \cite{PRK11} is achievable for any possible $d<n-1$ in the bandwidth-limited regime provided that the number of compromised nodes is below  a certain limit. Thus, we characterize the secure capacity of the DSS for the limited-knowledge adversary for the regime we consider. Our proof is constructive and is based on using the hashing schemes of \cite{PRK11} with an alternative representation of Product-Matrix codes \cite{RSK11}. Compared to the scheme of \cite{PRK11}, our scheme not only allows secure data reconstruction by the user, but also allows the regeneration of an uncorrupted exact copy of the lost data during repair from a failure.

\noindent\paragraph*{Organization} The paper is organized as follows. In Sec.~\ref{sec:Model}, we describe the system and adversary model. In Sec.~\ref{sec:main}, we state our main result. We follow  by first describing our secure  code construction in Sec.~\ref{sec:CodeConstruct} and an example on the decoding algorithm in Sec.~\ref{section:secure_decoding_ex}. The proof of the main result is given  in Sec.~\ref{sec:proof}.

\section{System Model}\label{sec:Model}
 An $(n,k,d)$ DSS  is formed  of $n$  unreliable nodes $\{1, 2, \ldots, n\}$, each with a storage capacity equal to $\alpha$ symbols. A DSS   allows any legitimate user  to reconstruct its stored   file  by connecting to any $k$ out of the $n$  nodes. Thus, it can tolerate $n-k$ simultaneous failures. We assume that the file symbols are the realizations of iid random variables uniformly distributed over a finite field.

  When a node fails, the DSS is repaired by  replacing the failed node with  a new one. The new node  connects to  $d, k \le d \le n-1,$ helper nodes,  chosen out of the remaining $n-1$ active ones, and downloads $\beta$ symbols from each.  These $\beta$ symbols are processed and the result is stored on the new node. So,  we have $\alpha\leq d\beta$.
 The literature distinguishes between two types of repair: {\em exact repair} where the regenerated data on the new node is an exact copy of the lost data, and {\em functional repair} where it is functionally equivalent in the sense that it retains the codes properties of file reconstruction and repair. Functional repair is more tractable for theoretical analysis, whereas practical systems typically require exact repair.

 {\em Adversary model:} Our objective is to guarantee  data reliability and integrity even when $b$ nodes in the DSS are compromised by an adversary. We focus on a {\em limited-knowledge adversary} that controls $b$ nodes and can read and corrupt all their stored data and/or their messages sent to other nodes during repair or to the users during file reconstruction. The only information the limited-adversary has about the stored data comes from the data on the nodes under its control. This is in contrast with an {\em omniscient adversary} that knows the data on all the nodes but controls only few nodes. We assume that the adversary has
complete knowledge of the storage and repair schemes implemented in
the DSS.

  {\em Trusted server:} In addition to the the previous setting, there is  a special node, referred to as a  trusted server\footnote{ In case the trusted server does not exist, it can be emulated using the nodes in the DSS using the technique in \cite{PRK11, JLKHKME08}.},  that can never be compromised by an adversary. The trusted server is considered to be an expensive resource and has a limited storage capacity. So, it cannot be used to store the data in the DSS. It will be used to store hashes of the data.

\section{Main result}\label{sec:main}
The secure capacity $C_s$ of a DSS against an adversary was defined in \cite{PRK11, PRKISIT11} to be the maximum amount of information that a user contacting $k$ nodes can always decode with an arbitrarily small probability of error and after any number of (functional) repairs. The following upper bound was proven in \cite{PRK11} in the case of  a limited-knowledge adversary,
\begin{equation}
C_s\leq \sum_{i=b+1}^{k} \min\left\{\alpha,(d-i+1)\beta\right\}.\label{eq:UpperBound}
\end{equation}

Setting $b=0$, i.e., when there is no adversary, we recover part of  the original result in \cite{DGWWR07} on the storage vs. repair bandwidth tradeoff for a file of size $B$,
\begin{equation}
B\leq \sum_{i=1}^{k} \min\left\{\alpha,(d-i+1)\beta\right\}.\label{eq:tradeoff}
\end{equation}
 Each term  in the summation in Eq.~\eqref{eq:tradeoff} accounts for the contribution of each of the $k$ nodes which is at most $\alpha$ symbols, but can be less due to correlation between the data on the different nodes. Therefore, an intuitive justification of Eq.~\ref{eq:UpperBound} is that one strategy of the adversary is to always delete the data on the nodes under its control (say always change it the all zero sequence). This results in the loss of the data on $b$ nodes, which in the worst case, can be among the $k$ nodes contacted by the user. It is worth mentioning that the effect of an omniscient adversary is the loss of $2b$ nodes from the DSS and the following upper bound was  the  shown in \cite{PRK11, PRKISIT11}, $C_s\leq \sum_{i=2b+1}^{k} \min\left\{\alpha,(d-i+1)\beta\right\}$ which can be regarded as a generalization of the Singleton bound. This bound was shown to be achievable in certain  regimes \cite{PRK11, PRKISIT11,RSK12}.

The only achievability result for the upper  bound in Eq.~\ref{eq:UpperBound} is known for $d=n-1$ in  the bandwidth-limited regime. In this regime, the only restriction is on the repair bandwidth $\beta$ and    there is no restriction on the storage capacity of a node $\alpha$, in particular $\alpha\geq d\beta$ (the new node can store all the downloaded data). Our main results stated in Th.~\ref{th:main} asserts the  achievability of this bound in the bandwidth-limited regime (including  the the so-called minimum-bandwidth  regime (MBR)) for any possible repair degree $d$ as long as $b< k/2$.

\begin{theorem}
The secure capacity of an $(n,k,d)$ DSS with exact repair in the bandwidth limited regime  is given by
\begin{equation}
C_s=\sum_{i=b+1}^{k} (d-i+1)\beta,
\end{equation}
when $b<k/2$ nodes are compromised by a limited knowledge-adversary.
\label{th:main}
\end{theorem}

The proof of Th.~\ref{th:main} is constructive and the rest of the paper will be dedicated to describing the scheme that achieves the secure capacity.

\section{Secure Code Construction}\label{sec:CodeConstruct}

Our construction is based  on  the  Product-Matrix (PM) codes introduced by Rashmi \textit{et al.} in \cite{RSK11} and the correlation hashing scheme in \cite{PRK11}. We  introduce a representation of  PM codes that is equivalent to the original codes in \cite{RSK11} in the sense that one can be obtained as an invertible linear transformation of the other. This representation   reflects  the intuition behind our secure scheme and simplifies the proof.

 \noindent{\em Original MBR PM code:}
 We consider an $(n,k,d)$ DSS. WLOG, we assume the repair bandwidth per link $\beta$ is normalized to $1$ ($\beta=1$). We choose $\alpha=d\beta=d$, which corresponds to the MBR regime.  We will explain the construction with an example by taking $(n,k,d) = (7,3,4)$ and following the notation in \cite{RSK11}. The maximum file size that can be stored is $B=9$ (see Eq.~\eqref{eq:tradeoff}). The encoding matrix is an $n\times d$ Vandermonde matrix $\Psi$ with elements  in $\mathbb{F}_q$. We denote the $i^{th}$ row of $\Psi$ by $\ubar{\psi}_i$. The stored file is denoted by $\ubar{U}=(U_1,...,U_B)$,  where  the symbols $U_i\in\mathbb{F}_{q^v}$ are packets of length $v$. The message symbols are arranged in an $d\times d$ matrix $M=\left[\begin{array}{cc}S&T\\T^t&0\end{array}\right]$, where, $S$ is a $k\times k$ symmetric matrix and $T$ is a $k\times(d-k)$ matrix and 0 is a $(d-k)\times (d-k)$ zero matrix. For our  example,
 \begin{equation*}\Psi = \left[\begin{array}{cccc}
     1  &   1  &   1  &  1\\
     1  &   2  &   4  &  8\\
     1  &   3  &   9  &  5\\
     1  &   4  &   5  &  9\\
     1  &   5  &   3  &  4\\
     1  &   6  &   3  &  7\\
     1  &   7  &   5  &  2\\
     \end{array}\right],\ M=\left[
                                  \begin{array}{cccc}
                                    U_1 & U_2 & U_3 & U_7 \\
                                    U_2 & U_4 & U_5 & U_8 \\
                                    U_3 & U_5 & U_6 & U_9 \\
                                    U_7 & U_8 & U_9 & 0 \\
                                  \end{array}
                                \right],
\end{equation*}
where  $\Psi$ is defined over $\mathbb{F}_{11}$. Node $i$  stores the coded vector $W_i=\ubar{\psi}_i M$. For example on node 1 we store $\ubar{\psi}_1 M=\left[\begin{array}{cccc}1&1&1&1\end{array}\right]M$. For details on repair and file reconstruction we refer the reader to \cite{RSK11}.

\noindent{\em Modified PM code representation:}

We define $X_{ij}\in\mathbb{F}_{q^v}$, $i,j\leq n,\ i\neq j$,  to be the symbol sent by node $i$ to node $j$ when node $j$ is being repaired. From \cite{RSK11}, we know that $X_{ij}=\ubar{\psi}_i M \ubar{\psi}_j^t$ (the superscript $t$ denotes the transpose operation). Note that  $X_{ij}=X_{ji}$ because $M$ is symmetric.
On each node $i$, we store any $\alpha=4$ different  symbols among $X_{i1},\dots, X_{i,j-1},X_{i,j+1}, \dots, X_{in}$. WLOG, we store the first $\alpha=4$ symbols
 as described in Table~\ref{data}. \begin{table}[h]
\centering
 \begin{tabular}{|c|c c c c|c|}
                      \hline
                      node 1 & $X_{12}$ & $X_{13}$ & $X_{14}$ & $X_{15}$ & $\{X_{16},X_{17}\}$ \\ \hline
                      node 2 & $X_{21}$ & $X_{23}$ & $X_{24}$ & $X_{25}$ & $\{X_{26},X_{27}\}$ \\ \hline
                      node 3 & $X_{31}$ & $X_{32}$ & $X_{34}$ & $X_{35}$ & $\{X_{36},X_{37}\}$ \\ \hline
                      node 4 & $X_{41}$ & $X_{42}$ & $X_{43}$ & $X_{45}$ & $\{X_{46},X_{47}\}$ \\ \hline
                      node 5 & $X_{51}$ & $X_{52}$ & $X_{53}$ & $X_{54}$ & $\{X_{56},X_{57}\}$ \\ \hline
                      node 6 & $X_{61}$ & $X_{62}$ & $X_{63}$ & $X_{64}$ & $\{X_{65},X_{67}\}$ \\ \hline
                      node 7 & $X_{71}$ & $X_{72}$ & $X_{73}$ & $X_{74}$ & $\{X_{75},X_{76}\}$ \\ \hline
                    \end{tabular}
\caption{\scriptsize{Symbols stored on each node in the modified PM code representation. The symbols between braces (\{\}) are not stored but can be computed using the stored symbols.}}\vspace{-0.6cm}
\label{data}
\end{table}

Here, the data stored on node $i$, say $W_i'$, can be written as $W_i'=W_iL$, with $L$ being an invertible matrix, hence the equivalence of the two representations. For instance, $W_1'=\ubar{\psi}_1M [\ubar{\psi}_2^t \,\, \ubar{\psi}_3^t\,\,  \ubar{\psi}_4^t\,\,  \ubar{\psi}_5^t]=W_1L$, where $L=[\ubar{\psi}_2^t \,\, \ubar{\psi}_3^t\,\,  \ubar{\psi}_4^t\,\,  \ubar{\psi}_5^t]$ is invertible because it is a submatrix of the Vandermonde matrix $\Psi$. Therefore, the exact repair and file reconstruction properties are directly inherited from the original PM codes.

 For each node $i$, any of  the symbols $X_{ij}$'s, can be computed from any $\alpha=4$ other symbols (follows directly from $\Psi$ being a Vandermonde matrix). Therefore, if node $1$ is being repaired by contacting helper nodes $2,3,4,5$, then it downloads $X_{12}, X_{13},X_{14}, X_{15}$ from each respectively. However, if it contacts say $2,3,4,6$, then it downloads $X_{12}, X_{13},X_{14}, X_{16}$ from each respectively. Then, computes $X_{15}$ using the downloaded data and stores $X_{12}, X_{13},X_{14}, X_{15}$.

\noindent{\em The hashes:}
The correlation hashing scheme introduced in \cite{PRK11} will be used  to cross-check the data on the different nodes against each other and will allow to detect the compromised nodes (with high probability) when used with  PM codes.
We  abuse  notation and write $X_{ij}$ as a vector in $\mathbb{F}_q^v$,  $X_{ij}=(X_{ij}^1,X_{ij}^2,\dots,X_{ij}^v)$, where $X_{ij}^k\in\mathbb{F}_q$. The hash of  two distinct symbols $X_{ij}$ and $X_{lk}$  is defined as the dot product  $X_{ij}.X_{lk}^t=\sum_{s=1}^{v}X_{ij}^sX_{lk}^s\in\mathbb{F}_q$. All these hashes are stored on the  trusted server and cannot be corrupted  by the adversary.

\section{Secure Decoding Algorithm: Example}\label{section:secure_decoding_ex}
Here, we describe how the hashing scheme in \cite{PRK11} can be used with the PM codes to achieve security.
We describe the secure repair and file reconstruction scheme through an example. Consider the problem of securing an  $(n,k,d)=(7,4,5)$ DSS against an  adversary controlling $b=1$ node(s).  We use the  $(n,k'=k-b,d'=d-b)=(7,3,4)$ PM code described in  Sec.~\ref{sec:CodeConstruct}. Note that, $k=4$ and $d=5$ nodes will still be contacted during file reconstruction and node repair respectively, i.e., $b=1$ extra node(s) more than what the actual PM code is designed for is contacted.  The idea is to use the hashes to identify the compromised node, then ignore its data by treating it as an erasure. The  back-off of  $b$ in $k$ and $d$ will allow to repair and reconstruct the data even in the presence of the erasure.  Therefore, we can guarantee the security of the data in the system and can detect and report the compromised nodes. In the following discussion, we focus on the detection of the corrupted nodes. Notice that the limited-knowledge adversary controlling $b=1$ node(s), say node $3$, can observe the data stored on this node and only $b=1$ symbol(s) on each other node. This happens because $X_{3i}=X_{i3}$ for all $i=1,\dots, n$.

\subsection{Secure repair}
Assume WLOG that in Table~\ref{data} node 3 is compromised and that node 2 fails. The new node contacts $d=5$ nodes, say nodes  $1,3,4,5$ and $6$ for repair. Each helper node $h$ sends $X_{h2}=X_{2h}$ to the new node $2$. The new node computes the correlation hashes of the downloaded data and compares it to the corresponding hashes downloaded from the trusted server. Then, it constructs the symmetric  comparison  table similar to the example  shown in table \ref{table:repair}.

\begin{table}[h]
\centering
\begin{tabular}{|c|c|c|c|c|c|}
  \hline
  ~ & $X_{12}$ & $X_{32}$ & $X_{42}$ & $X_{52}$ & $X_{62}$ \\ \hline
  $X_{12}$ & ~ & $\times$ & $\checkmark$ & $\checkmark$ & $\checkmark$ \\ \hline
  $X_{32}$ & $\times$ & ~ & $\times$ & $\times$ & $\times$ \\ \hline
  $X_{42}$ & $\checkmark$ & $\times$ & ~ & $\checkmark$ & $\checkmark$  \\ \hline
  $X_{52}$ & $\checkmark$ & $\times$ & $\checkmark$ & ~ & $\checkmark$  \\ \hline
  $X_{62}$ & $\checkmark$ & $\times$ & $\checkmark$ & $\checkmark$ & ~  \\
  \hline
\end{tabular}\\
\caption{\scriptsize{ A comparison of the computed hashes of the  downloaded data vs. the downloaded hashes from the trusted server. A blank indicates that the product is not computed, a $\checkmark$ indicates that the hashes are equal, while an $\times$ indicates that the hashes do not match.}}\vspace{-.6cm}
\label{table:repair}
\end{table}

The repaired node looks for a column with $2$ $\times$'s or more to identify the compromised node. The reason is that a column with a single $\times$, say in the entry corresponding to $X_{12}$ and  $X_{32}$, is confusing and can mean either node $2$ or node $3$ are   compromised.

 The example in Table~\ref{table:repair} assumes the ideal case in which all the corrupted symbols result in a hash mismatch. However,  the adversary can always attempt to corrupt the data in a way to match the downloaded hashes, i.e,  create a hash collision. For example, it will be able to fool the new node if  it manages to change at least $3$ $\times$'s into $\checkmark$'s in the column of $X_{32}$.  However, its observation consisting of only $(X_{31}, X_{32}, X_{34}, X_{35})$ here, is uncorrelated (linearly independent) with the other downloaded packets  $X_{12}, X_{42}, X_{52}$ (follows from the construction of PM codes). And, its best strategy cannot be better than introducing random errors which could be caught with high probability (see Section~\ref{subsection:error_analysis}).
  After identifying the compromised node(s), the new node will regenerate the lost   data  using the symbols received from the other helper nodes and report the compromised node(s).

\subsection{Secure reconstruction}

To reconstruct a file, a user contacts any  $k=4$ nodes, say nodes  $1, 2, 3$ and $4$, downloads all their symbols and computes their correlation hashes. Then, the user compares the computed hashes with their corresponding ones downloaded  from the trusted server and constructs  a $k\times k=4\times 4$ block-table formed of blocks each of size $(d-b)\times (d-b)=4\times4$. Table~\ref{table:reconstable} depicts part of such table corresponding to nodes $3$ and $4$.

We say that a block is {\em mismatched} if it has at least one $\times$. In the reconstruction block-table, the column corresponding to a non-compromised node will contain at most $b=1$ mismatched block(s). Therefore, the user will identify the compromised node by looking for a column with more than one mismatched block. The user then discards the data from the compromised node and decode the file using the symbols downloaded from the other $3$ nodes. It can do so because the underlying PM code has an effective $k'=3$. Here too, the adversary can attempt to avoid being detected by choosing  erroneous symbols that lead to a match between the computed and trusted hashes. For instance here, the adversary needs to flip all the $\times$'s in two blocks in the column of node $3$ in order to be undetected. Next, we will analyze the probability of this event and show that it can be made arbitrarily small.

\begin{table}[h]
\centering
\begin{tabular}{|c|c||c|c|c|c||c|c|c|c|}
  \cline{3-10}
  \multicolumn{2}{c|}{~} &\multicolumn{4}{c||}{node 3}&\multicolumn{4}{c|}{node 4}\\ \cline{3-10}
  \multicolumn{2}{c|}{~}& $X_{31}$ & $ X_{32}$ & $X_{34}$ & $X_{35}$& $X_{41}$ & $ X_{42}$ & $X_{43}$ & $X_{45}$\\ \cline{3-10} \hline
  \parbox[t]{2mm}{\multirow{4}{*}{\rotatebox[origin=c]{90}{node1}}}
  & $X_{12}$  & $\times$ & $\times$ & $\times$ & $\times$ & $\checkmark$ & $\checkmark$ & $\checkmark$ & $\checkmark$\\ \cline{2-10}
  & $X_{13}$ & $\checkmark$ & $\checkmark$ & $\checkmark$ & $\checkmark$ & $\checkmark$ & $\checkmark$ & $\checkmark$ & $\checkmark$\\ \cline{2-10}
  & $X_{14}$ & $\times$ & $\times$ & $\times$ & $\times$  & $\checkmark$ & $\checkmark$ & $\checkmark$ & $\checkmark$\\ \cline{2-10}
  & $X_{15}$ & $\times$ & $\times$ & $\times$ & $\times$  & $\checkmark$ & $\checkmark$ & $\checkmark$ & $\checkmark$\\ \hline \hline
  \parbox[t]{2mm}{\multirow{4}{*}{\rotatebox[origin=c]{90}{node 2}}}
  & $X_{21}$ &$\times$ & $\times$ & $\times$ & $\times$   & $\checkmark$ & $\checkmark$ & $\checkmark$ & $\checkmark$\\ \cline{2-10}
  & $X_{23}$ & $\checkmark$ & $\checkmark$ & $\checkmark$ & $\checkmark$ & $\checkmark$ & $\checkmark$ & $\checkmark$ & $\checkmark$\\ \cline{2-10}
  & $X_{24}$ & $\times$ & $\times$ & $\times$ & $\times$  & $\checkmark$ & $\checkmark$ & $\checkmark$ & $\checkmark$\\ \cline{2-10}
  & $X_{25}$ & $\times$ & $\times$ & $\times$ & $\times$  & $\checkmark$ & $\checkmark$ & $\checkmark$ & $\checkmark$\\ \hline \hline
  \parbox[t]{2mm}{\multirow{4}{*}{\rotatebox[origin=c]{90}{node 3}}}
  & $X_{31}$ & ~ & ~ & ~ & ~& $\times$ & $\times$ & $\checkmark$ & $\times$\\ \cline{2-10}
  & $X_{32}$ & ~ & ~ & ~ & ~& $\times$ & $\times$ & $\checkmark$ & $\times$\\ \cline{2-10}
  & $X_{34}$ & ~ & ~ & ~ & ~& $\times$ & $\times$ & $\checkmark$ & $\times$\\ \cline{2-10}
  & $X_{35}$ & ~ & ~ & ~ & ~& $\times$ & $\times$ & $\checkmark$ & $\times$\\ \hline \hline
  \parbox[t]{2mm}{\multirow{4}{*}{\rotatebox[origin=c]{90}{node 4}}}
  & $X_{41}$ & $\times$ & $\times$ & $\times$ & $\times$& ~ & ~ & ~ & ~\\ \cline{2-10}
  & $X_{42}$ & $\times$ & $\times$ & $\times$ & $\times$& ~ & ~ & ~ & ~\\ \cline{2-10}
  & $X_{43}$ & $\checkmark$ & $\checkmark$ & $\checkmark$ & $\checkmark$ & ~ & ~ & ~ & ~\\ \cline{2-10}
  & $X_{45}$ & $\times$ & $\times$ & $\times$ & $\times$& ~ & ~ & ~ & ~\\ \hline
\end{tabular}
\caption{\scriptsize{Part of the reconstruction hash table formed by a user contacting nodes 1, 2, 3 and 4. The columns corresponding to nodes 1 and 2 are omitted here due to space constraints.}}\vspace{-.8cm}\label{table:reconstable}
\end{table}

\subsection{Error analysis}\label{subsection:error_analysis}
The error analysis  in \cite{PRK11} can be applied here to prove that the probability of failing to detect a compromised node  during repair or  reconstruction is upper bounded by  $1/q$ and therefore can be made as small as desired by increasing $q$. Suppose the adversary wants to  flip an $\times$ to a $\checkmark$ in a comparison table, say the $\times$ corresponding to $X_{12}$ and $X_{32}$ in Table~\ref{table:repair}. It can introduce an error $e$ to $X_{32}$ as to change it to $X_{32}+e$ and will succeed in flipping the entry in the table if $e$ is orthogonal to $X_{12}$. Assuming the symbols of the original file are iid and uniformly distributed over $\mathbb{F}_q^v$,
for a fixed error $e \in \ \mathbb{F}_q^v$, there are  $q^{v-1}$ possible values of $X_{12}$ that  are orthogonal to $e$. Hence, the probability of $e$ being orthogonal to one symbol given the adversary's observation is
\begin{equation*}
P\left(e\bot X_{12}\right)=\frac{q^{v-1}}{q^v}=\frac{1}{q}.
\end{equation*}
Of course, if the adversary were omniscient it would  know the value of  $X_{12}$ and  can  choose a value of $e$ that is orthogonal to $X_{12}$. Note that the adversary needs to flip many entries in the comparison table to fool the secure repair and/or the reconstruction algorithms described earlier. These events are not independent and we upper bound the probability of error, i.e., adversary being undetected, by $1/q$.

\section{Proof of Main Result}\label{sec:proof}

We  show   that the code construction described in Sec.~\ref{sec:CodeConstruct} based on PM codes and correlation hashes achieves the resiliency capacity in Th.~\ref{th:main}.
To secure an $(n,k,d)$ DSS against an adversary controlling $b$ nodes, the construction uses  an $(n,k'=k-b,d'=d-b)$ minimum-bandwidth regenerating PM  code with storage per node $\alpha=(d-b)\beta$. These codes achieve the (non-secure) optimal storage/repair bandwidth tradeoff  and can store a file of size $B=\sum_{i=1}^{k-b}\min{\left\{\alpha,(d-b-i+1)\beta\right\}}$ symbols \cite{DGWWR07, RSK11}. A simple change of variables shows that $B=C_s$.  We prove  Th.~\ref{th:main} by showing that these $B$ symbols can be secured by following the secure repair and secure decoding rules in Sec~\ref{section:secure_decoding_ex}.

\subsection{Secure repair}
If a node $f$ fails, the new node contacts $d$ helper nodes, $h_1,h_2,\dots,h_d$ for repair. Each of the helper nodes send $X_{h_if}$ to the new node. In addition, the new node downloads all the hashes $X_{h_if}\cdot X_{h_jf}$  from the trusted server and forms a repair comparison table similar to Table~\ref{table:repair}.
WLOG, suppose  that among the helper nodes  exactly $b$  nodes are compromised\footnote{The analysis stays the same even if less  than $b$ helper nodes happen to be compromised.}.  Also, suppose the ideal case that  every  transmitted packet  corrupted by the adversary results in an $\times$ in the comparison table. An uncorrupted node will contain at most $b$ $\times$'s in its column.  Therefore, every helper node corresponding to more than $b$ $\times$'s must be a compromised node.

{\bf Secure repair  rule:} {\em The repaired node  decides that any helper node corresponding to a column (or a row) in the comparison table with more than  $b$ $\times$'s is a compromised node.}

  The adversary can try to avoid being detected by causing errors that lead to switching at least $d-b$ $\times$'s into $\checkmark$'s in columns corresponding to nodes under its control. However,
  since we assume that $b<k/2$ the adversary knows zero information about the symbols sent by the helpers node that are not under its control (a property of the underlying $(n,k-b,d-b)$ PM code). Therefore, the error analysis detailed in Sec.~\ref{subsection:error_analysis} holds here and the probability of the adversary succeeding, i.e., causing undetected erroneous repair data, is upper bounded by $1/q$.

\subsection{Secure reconstruction}

A user that  wants to reconstruct the file contacts $k$ nodes and downloads all their stored data ($\alpha=d-b$ symbols from each node). In addition, it contacts the trusted server and downloads all the hashes corresponding to the downloaded symbols. WLOG, suppose that $b$ nodes among those which were contacted are compromised. Then, the user forms the reconstruction block-table between the trusted hashes and computed hashes as done in Table~\ref{table:reconstable}.

Each block contains the hashes cross-checking the symbols downloaded from two different nodes.  A mismatched block indicates that one of these two  nodes is sending corrupted packets and therefore is  compromised.
 Hereafter,  a node having more than $b+1$ mismatched blocks has to be compromised. Otherwise, the $b+1$ other nodes corresponding to the mismatched blocks are all compromised which contradicts our initial assumption that there are at most $b$ compromised nodes. Therefore, the user implements the following rule:

{\bf Secure reconstruction rule:} {\em The user decides that any node having more than $b$ mismatched blocks in its row or column is a compromised node. }

Treating the $b$ compromised nodes as erasures, the user recovers the data using the $k-b$ other contacted nodes. Moreover, it  reports the identity of the compromised nodes.

Note that if $b\geq k/2$, then the  limited-knowledge adversary can decode the whole file and is actually omniscient.  In this case, our scheme cannot be used to achieve security since the adversary can carefully choose the errors as to switch all the $\times$'s into $\checkmark$'s in the comparison tables. If $b< k/2$, then on each uncompromised node there is at least one symbol that the adversary does not have information about, and the error analysis in Sec.~\ref{subsection:error_analysis} holds again here.

\subsection{Hash storage overhead}
The total number of stored hashes is $\binom{\theta}{2}$ symbols in $\mathbb{F}_q$, where $\theta=\binom{n}{2}$ is the total number of the  $X_{ij}$ symbols. The ratio of the hash size to the stored data size is $\frac{\binom{\theta}{2}}{n \alpha v}=\mathcal{O}(\frac{1}{v})$ which can be made arbitrarily small by increasing the packet length $v$. In the given example this ratio is $\frac{210}{7\times 4 v} = \frac{15}{2v}$.

\bibliographystyle{ieeetr}
\bibliography{DSS}

\end{document}